\documentclass[aps,prc,twocolumn,showpacs,superscriptaddress,amsmath,amssymb,showkeys]{revtex4}
\usepackage{graphicx}
\usepackage{dcolumn}
\usepackage{bm}

\begin{document}

\title{Upper limit of the total cross section for the $pn \to pn\eta^{\prime}$ reaction}
\author{J.~Klaja}
\affiliation{Institute of Physics, Jagiellonian University, PL-30-059 Cracow, Poland}
\affiliation{Institute for Nuclear Physics and J{\"u}lich Center for Hadron Physics,\\
             Research Center J{\"u}lich, D-52425 J{\"u}lich, Germany}
\author{P.~Moskal} \email[Electronic address: ]{p.moskal@fz-juelich.de}
\affiliation{Institute of Physics, Jagiellonian University, PL-30-059 Cracow, Poland}
\affiliation{Institute for Nuclear Physics and J{\"u}lich Center for Hadron Physics,\\
             Research Center J{\"u}lich, D-52425 J{\"u}lich, Germany}
\author{S.~D.~Bass}
\affiliation{Institute for Theoretical Physics, University of Innsbruck, Austria}
\author{E.~Czerwi\'nski}
\affiliation{Institute of Physics, Jagiellonian University, PL-30-059 Cracow, Poland}
\affiliation{Institute for Nuclear Physics and J{\"u}lich Center for Hadron Physics,\\
             Research Center J{\"u}lich, D-52425 J{\"u}lich, Germany}
\author{R.~Czy\.zykiewicz}
\affiliation{Institute of Physics, Jagiellonian University, PL-30-059 Cracow, Poland}
\author{D.~Gil}
\affiliation{Institute of Physics, Jagiellonian University, PL-30-059 Cracow, Poland}
\author{D.~Grzonka}
\affiliation{Institute for Nuclear Physics and J{\"u}lich Center for Hadron Physics,\\
              Research Center J{\"u}lich, D-52425 J{\"u}lich, Germany}
\author{T.~Johansson}
\affiliation{Department of Physics and Astronomy, Uppsala University, Sweden}
\author{B.~Kamys}
\affiliation{Institute of Physics, Jagiellonian University, PL-30-059 Cracow, Poland}
\author{A.~Khoukaz}
\affiliation{IKP, Westf\"alische Wilhelms-Universit\"at, D-48149 M\"unster, Germany}
\author{P.~Klaja}
\affiliation{Institute for Nuclear Physics and J{\"u}lich Center for Hadron Physics,\\
             Research Center J{\"u}lich, D-52425 J{\"u}lich, Germany}
\affiliation{Physikalisches Institut, Universit{\"a}t Erlangen-N{\"u}rnberg, D-91058 Erlangen, Germany}
\author{W.~Krzemie{\'n}}
\affiliation{Institute of Physics, Jagiellonian University, PL-30-059 Cracow, Poland}
\affiliation{Institute for Nuclear Physics and J{\"u}lich Center for Hadron Physics,\\
             Research Center J{\"u}lich, D-52425 J{\"u}lich, Germany}
\author{W.~Oelert}
\affiliation{Institute for Nuclear Physics and J{\"u}lich Center for Hadron Physics,\\
             Research Center J{\"u}lich, D-52425 J{\"u}lich, Germany}
\author{B.~Rejdych}
\affiliation{Institute of Physics, Jagiellonian University, PL-30-059 Cracow, Poland}
\author{J.~Ritman}
\affiliation{Institute for Nuclear Physics and J{\"u}lich Center for Hadron Physics,\\
             Research Center J{\"u}lich, D-52425 J{\"u}lich, Germany}
\author{T.~Sefzick}
\affiliation{Institute for Nuclear Physics and J{\"u}lich Center for Hadron Physics,\\
             Research Center J{\"u}lich, D-52425 J{\"u}lich, Germany}
\author{M.~Siemaszko}
\affiliation{Institute of Physics, University of Silesia, PL-40-007 Katowice, Poland}
\author{M.~Silarski}
\affiliation{Institute of Physics, Jagiellonian University, PL-30-059 Cracow, Poland}
\author{J.~Smyrski}
\affiliation{Institute of Physics, Jagiellonian University, PL-30-059 Cracow, Poland}
\author{A.~T\"aschner}
\affiliation{IKP, Westf\"alische Wilhelms-Universit\"at, D-48149 M\"unster, Germany}
\author{M.~Wolke}
\affiliation{Institute for Nuclear Physics and J{\"u}lich Center for Hadron Physics,\\
                 Research Center J{\"u}lich, D-52425 J{\"u}lich, Germany}
\affiliation{Department of Physics and Astronomy, Uppsala University, Sweden}
\author{P.~W\"ustner}
\affiliation{Institute for Nuclear Physics and J{\"u}lich Center for Hadron Physics,\\
                 Research Center J{\"u}lich, D-52425 J{\"u}lich, Germany}
\author{J.~Zdebik}
\affiliation{Institute of Physics, Jagiellonian University, PL-30-059 Cracow, Poland}
\author{M.~Zieli\'nski}
\affiliation{Institute of Physics, Jagiellonian University, PL-30-059 Cracow, Poland}
\author{W.~Zipper}
\affiliation{Institute of Physics, University of Silesia, PL-40-007 Katowice, Poland}
\date{\today}
             
\begin{abstract}
The upper limit of the total cross section for the $pn \to pn\eta^{\prime}$
reaction has been determined near the kinematical threshold in the excess energy 
range from 0 to 24~MeV. The measurement was performed using the COSY-11 detector 
setup, a deuteron cluster target, and the proton beam of COSY with a momentum of 3.35~GeV/c. 
The energy dependence of the upper limit of the cross section was extracted exploiting 
the Fermi momenta of nucleons inside the deuteron. Comparison of the determined upper 
limit of the ratio $R_{\eta^{\prime}}~=~{{\sigma(pn \to pn\eta^{\prime})} / {\sigma(pp \to pp\eta^{\prime})}}$
with the corresponding ratio for $\eta$-meson production does not favor the dominance of the $N^*(1535)$ 
resonance in the production process of the $\eta^{\prime}$ meson 
and suggests nonidentical production mechanisms for $\eta$ and $\eta^{\prime}$ mesons.

\vspace{1pc}
\end{abstract}
\maketitle

\section{Introduction}

Studies of $\eta$ and $\eta^{\prime}$ mesons and their interactions with 
nucleons provide an interesting window on axial U(1) dynamics.
The flavor-singlet $J^P = 1^+$ channel
is characterized by a large Okubo-Zweig-Iizuka (OZI) violation:
the masses of the $\eta$ and $\eta'$ mesons are 300--400 MeV larger
than the values they would have if these mesons were pure would-be 
Goldstone bosons associated with spontaneously broken chiral 
symmetry~\cite{bass-pst99,bass-app11}.

One needs extra mass in the singlet channel associated with
nonperturbative topological gluon configurations and
the QCD axial anomaly \cite{shore-lnp737}.
The strange quark mass induces considerable $\eta$-$\eta'$ mixing.
For the physical $\eta$-$\eta'$ mixing angle $-18$ degrees,
the singlet component in the $\eta$ has the potential to induce a factor
of 2 increase in the $\eta$-nucleon scattering length $a_{\eta N}$
relative to the value one would expect if the $\eta$ were a pure octet
state \cite{bass-plb634}.
The flavor-singlet Goldberger-Treiman relation 
\cite{tgv} relates  $\eta^{\prime}$-nucleon coupling to 
the small flavor-singlet axial charge measured in the proton spin puzzle
\cite{bass-rmp}.

The properties of $\eta$ and $\eta^{\prime}$ mesons should  manifest 
themselves in production and decay processes and in their interaction 
with nuclear matter.
These processes are being studied in experiments from threshold~\cite{moskal-hab}
through  high-energy collisions where anomalously high branching ratios have been
observed for $D_s$ and $B$-meson decays to an $\eta'$ plus additional 
hadrons \cite{ball-plb365,fritzsch-plb415}.

At the hadronic level, it is expected that the
$\eta^{\prime}$ meson can be produced through the exchange and 
rescattering of mesons, 
through the excitation of
an intermediate baryonic resonance 
or via the fusion of virtual mesons
~\cite{nakayama-prc61,kampfer-ep,cao-prc78,bass-plb463,nakayama-prc69}.
So far the $pn\to pn\eta^{\prime}$ process has been studied quantitatively
only within 
(i) an effective Lagrangian approach, 
assuming that the S$_{11}$(1535) resonance is dominant~\cite{cao-prc78}
and (ii) within  a covariant effective meson-nucleon 
theory including meson and nucleon currents with 
the nucleon resonances S$_{11}$(1650), P$_{11}$(1710) and P$_{13}$(1720)
\cite{kampfer-ep}. 
However, for $\eta'$ production there is an additional potential contribution
where glue is excited in the 
``short distance'' ($\sim 0.2$-fm)
interaction region of
the proton-nucleon collision and 
then evolves to become an $\eta'$ in the final state.
This gluonic contribution to the cross section 
for $pp \rightarrow pp \eta'$, proposed by Bass \cite{bass-plb463}, 
is beyond  the contributions associated with 
meson exchange models. 
There is no reason, a priori, to expect this OZI violating process
to be small.
Because glue is flavor-blind it contributes with the same strength 
in both 
the $pp \rightarrow pp \eta'$ and the $pn \rightarrow pn \eta'$ reactions.

It is not possible to establish the relative contributions of possible 
reaction mechanisms
responsible for $\eta^{\prime}$-meson production 
based only on the presently available cross sections 
for the $pp \to pp\eta^{\prime}$ reaction~\cite{daneppetaprim}.
Therefore, one has to investigate  $\eta^{\prime}$ production 
as a function of spin and isospin degrees of freedom.
Such studies were conducted in the case of the $\eta$ meson~\cite{calen-prc58,moskal-prc79,rafalprl}.
The ratio
$R_{\eta} 
 = \sigma (pn \rightarrow pn \eta ) / \sigma (pp \rightarrow pp \eta )$
has been measured for quasifree $\eta$ 
production from a deuteron target up to 109 MeV above threshold
\cite{calen-prc58,moskal-prc79}.
One finds that $R_{\eta}$ 
is approximately energy independent with a value of
$\sim6.5$ 
in the energy range of $16 - 109$ MeV
signifying a strong isovector exchange 
contribution to the $\eta$ production mechanism.

The ratio $R_{\eta^\prime}$ has not been measured to date,
and the existing predictions differ drastically 
depending on the model.
Cao and Lee~\cite{cao-prc78}
assumed, 
by analogy to the production of the $\eta$ meson,
that the production of 
the $\eta^{\prime}$ meson proceeds 
dominantly via the S$_{11}$(1535) resonance.
As a consequence they predicted within an effective Lagrangian approach 
an $R_{\eta^{\prime}}$ value
equal to the experimentally established $R_{\eta}$ value.
In contrast, 
Kaptari and K\"ampfer~\cite{kampfer-ep} 
predicted  a value of $R_{\eta^{\prime}}$
close to $\sim$1.5 in the kinematic range of the COSY-11 experiment 
with the dominant contribution coming from the  meson conversion currents. 
In the extreme scenario of glue-induced production saturating 
the $\eta^{\prime}$ production cross section, the ratio
$R_{\eta'}$
would approach unity
after correction for the final-state interaction between the two outgoing 
nucleons.

The SU(3) wave functions of the $\eta$ and $\eta^{\prime}$ mesons differ due 
to iso-singlet gluonic and strangeness degrees of freedom.
Hence, it is reasonable that singlet currents will
play a 
greater role in $\eta^{\prime}$  than in $\eta$ production.
Proton-proton data are available from COSY and SATURNE~\cite{daneppetaprim}.
To investigate the production mechanism
it is important to provide an empirical base with spin and 
isospin dependence of the cross sections.
In this article we  present first results 
on the measurement of the $pn\to pn\eta^{\prime}$ reaction.

\section{Experimental technique}

\begin{figure}[h]
\centering 
\includegraphics[width=0.35\textwidth,angle=0]{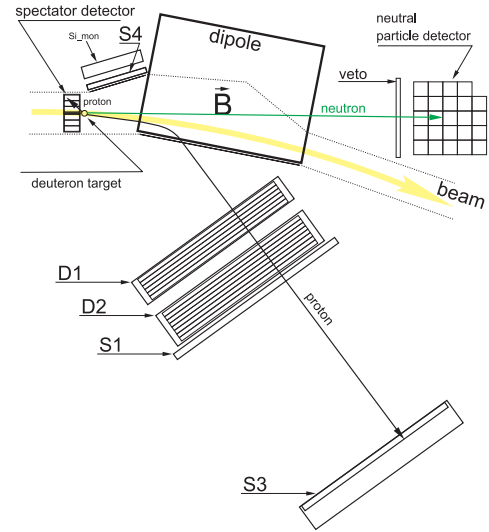}
\caption{Scheme of the COSY-11 detector system with superimposed
          tracks from the $pd \to p_{spec}(pn\eta^{\prime})$ reaction.
          Protons are registered in two drift chambers D1 and D2
          and in the scintillator hodoscopes S1 and S3. The S1 detector is
          built of 16 vertically arranged modules, whereas the S3 hodoscope
          is a nonsegmented scintillator wall viewed by a matrix
          of 217 photomultipliers.
          An array of silicon pad detectors (spectator~detector)
          is used for registration of spectator protons.
          Neutrons are registered in the neutral particle detector
          consisting of 24 independent detection units.
          To distinguish neutrons from charged particles a veto
          detector is used. Elastically scattered protons are measured
          in the scintillator detector S4 and the position-sensitive silicon 
          detector
          $Si_{mon}$. The sizes of the detectors and their relative distances
          are not to scale. }
\label{cosy11}
\end{figure}

The COSY-11 detector system~\cite{brauksiepe-nim} is schematically depicted 
in Fig.\ref{cosy11}.
A quasifree proton-neutron reaction was induced by a proton beam~\cite{cosy} with 
a momentum of 3.35~GeV/c impinging on a deuteron cluster target~\cite{target}. 
A detailed description  of the functioning of the detectors and
of the experimental technique is given in a recent COSY-11 
article devoted to measurement of 
the $pn\to pn\eta$ reaction~\cite{moskal-prc79} and in Ref.~\cite{jklaja-phd}.
Therefore,  we describe it only briefly here.

\begin{figure}[h]
\centering 
\includegraphics[width=0.33\textwidth,angle=0]{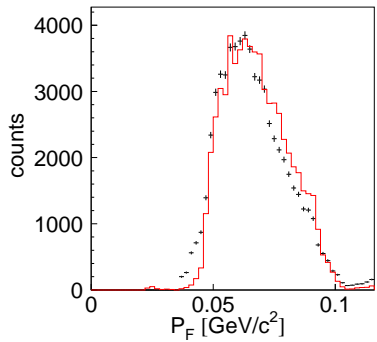}
\caption{
         Momentum distribution of
         the spectator proton as reconstructed in the experiment (crosses)
         in comparison with simulations taking into account
         the Fermi momentum distribution of nucleons inside the deuteron
         according to the Paris potential~\cite{lacombe-plb101}
         (solid histogram). Simulated events were analyzed in the same
         way as the experimental data. }
\label{banan_all_pads}
\end{figure}

The experiment is based on the registration of all outgoing nucleons
from the $pd\to ppnX$ reaction.
The proton moving forward is measured in two drift
chambers and scintillator detectors and the neutron is registered
in the neutral particle detector.
The proton moving backward is measured 
with a dedicated
silicon-pad detector~\cite{bilger-nim457}. 
In the data analysis
the backward proton from the deuteron is considered as a spectator that does 
not interact with the bombarding proton, but escapes untouched and hits the detector
carrying the Fermi momentum possessed at the time of the reaction.
The total energy available for the quasifree proton-neutron
reaction is calculated for each event based on the momentum vectors
of the spectator and beam protons
under the assumption that  the spectator proton 
is on its mass shell at the moment of the collision.
The absolute momentum of the neutron is determined 
from the time of flight between the target and the neutron detector and 
its direction is derived from the hit position
defined as the center of the detection module whose signal was produced 
as the first one.
The veto detector installed in front of the neutral particle detector 
discriminates signals originating from charged particles, 
whereas the $\gamma$ quanta are discernd by means of the time of flight measured
between the target and the neutral particle detector where the time of the 
reaction in the target is reconstructed based on the trajectory and 
velocity of the forward scattered
protons detected in drift chambers and S1 and S3 scintillators.  
Figure~\ref{banan_all_pads} presents the momentum distribution
of protons considered as spectators (crosses) compared with simulations
taking into account a Fermi motion of nucleons inside the deuteron 
(solid line).
In the analysis of the $pn\to pn\eta^{\prime}$ reaction
spectator protons with momenta ranging from about 35
to about 115~MeV/c were taken into account. 
The lower limit is caused by the noise level of the silicon pad detectors.
The {\it spectator noise cut} was determined for each pad of the 
detector separately, using the spectra of energy loss triggered by a 
pulser with a frequency of 1~Hz in addition to other experimental triggers.
Protons with a momentum larger than  about 115~MeV/c were not taken into account 
because we used the second layer of the double-layer spectator detector as a veto
to reduce the background from charged pions.
The shapes of the simulated and experimental spectra shown in 
Fig.~\ref{banan_all_pads}
agree quite well, and the small difference will be taken for 
the estimation of the systematical uncertainties of the final 
results caused by the assumption of a Fermi momentum distribution 
of nucleons inside the deuteron.

The accuracy of the excess energy determination was estimated based on 
Monte Carlo simulations. 
For this purpose we simulated $10^9$ events of 
the $pn\to pn\eta^{\prime}$ reaction taking
into account the Fermi motion of the nucleons inside the deuteron,
the size of the target~\cite{target}, 
the spread of the beam momentum
and the horizontal 
and vertical 
beam size~\cite{moskal-nim466}.
Simulated data were analyzed in the same way as the experimental 
data, taking into consideration 
the energy and time resolution
of the detectors, 
resulting in an excess energy resolution of $\sigma(Q)~=~4.2$~MeV.
It is worth stressing that using this method
for the $pn \to pn\eta$ reaction the standard deviation of
the excess energy was derived to be 
$\sigma(Q)~=~2.2$~MeV~\cite{moskal-prc79},
which is comparable to the $\sigma(Q)~=~1.8$~MeV obtained 
under similar conditions
at the PROMICE/WASA setup~\cite{bilger-nim457}.

\begin{figure}[h]
\centering 
\includegraphics[width=0.27\textwidth,angle=0]{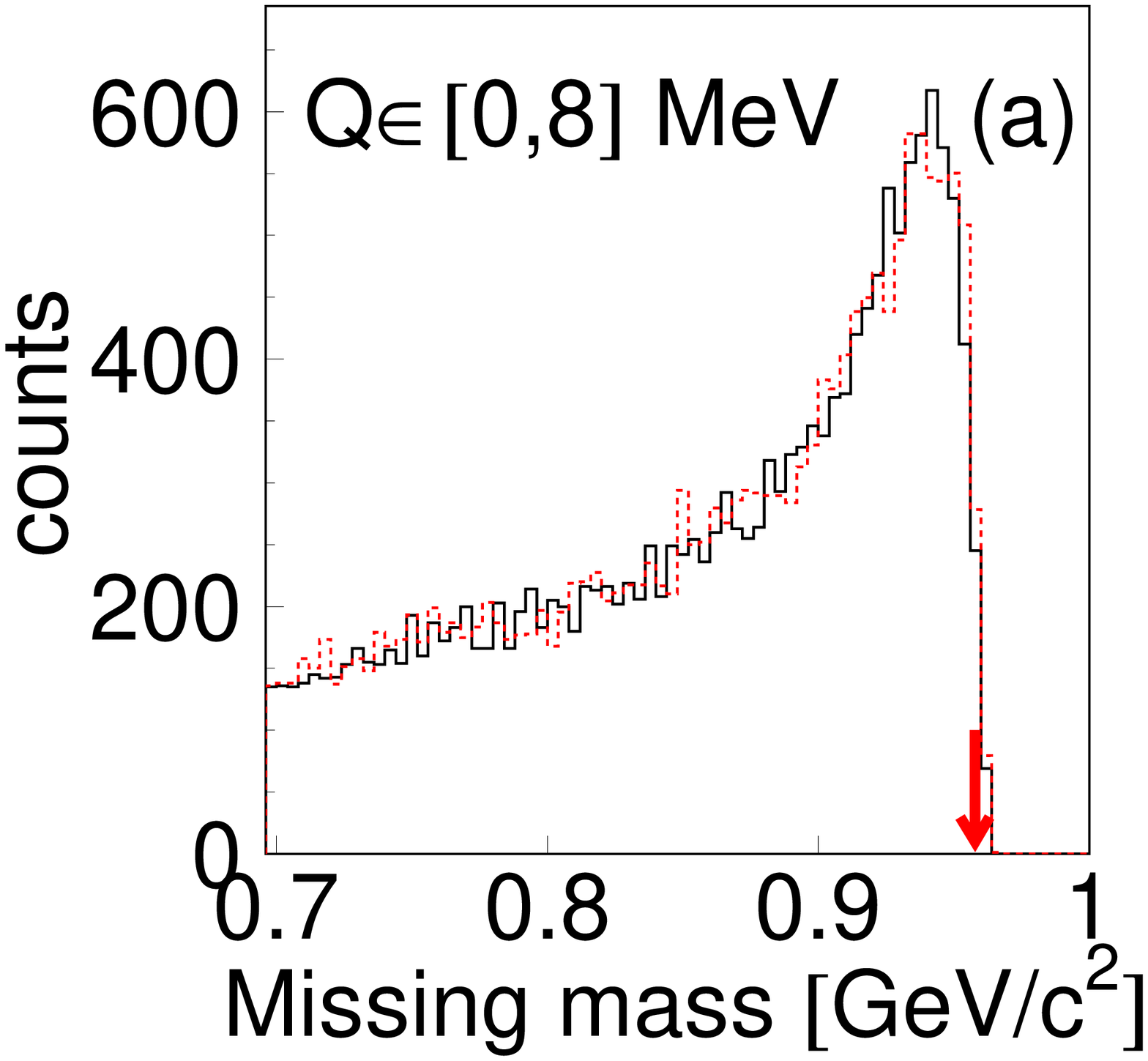}
\hspace{-1.25cm}
\includegraphics[width=0.27\textwidth,angle=0]{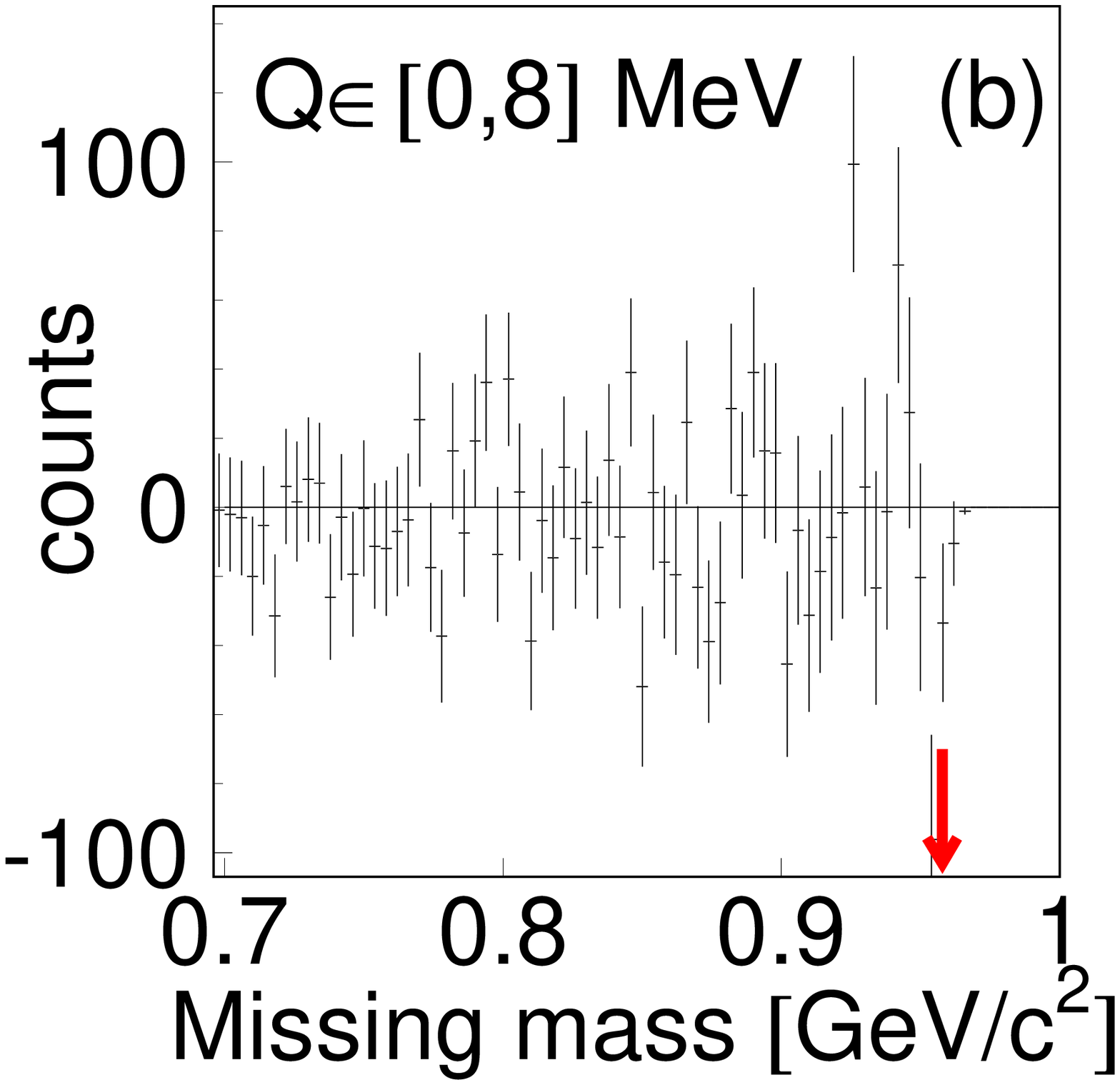}
\includegraphics[width=0.27\textwidth,angle=0]{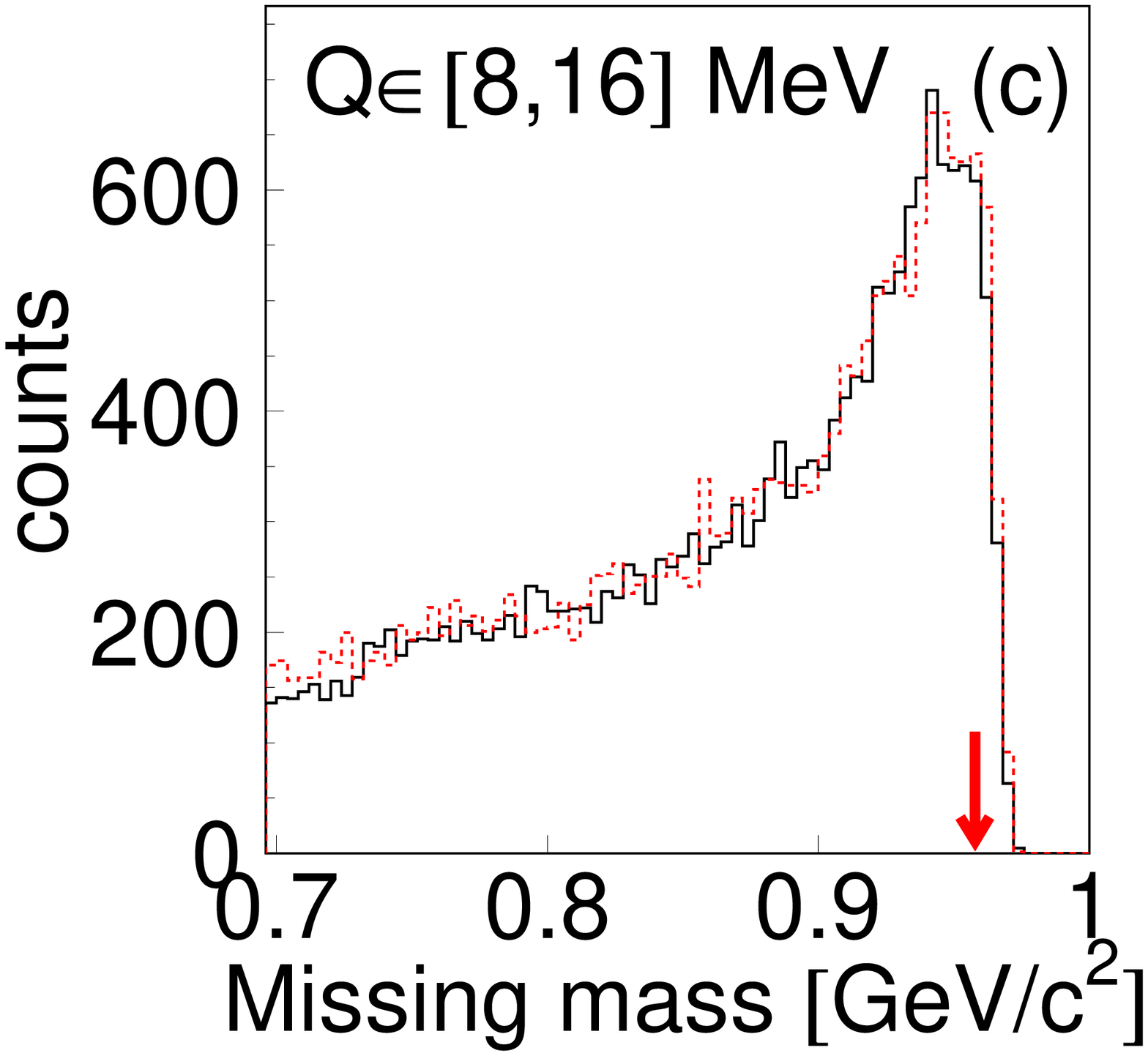}
\hspace{-1.25cm}
\includegraphics[width=0.27\textwidth,angle=0]{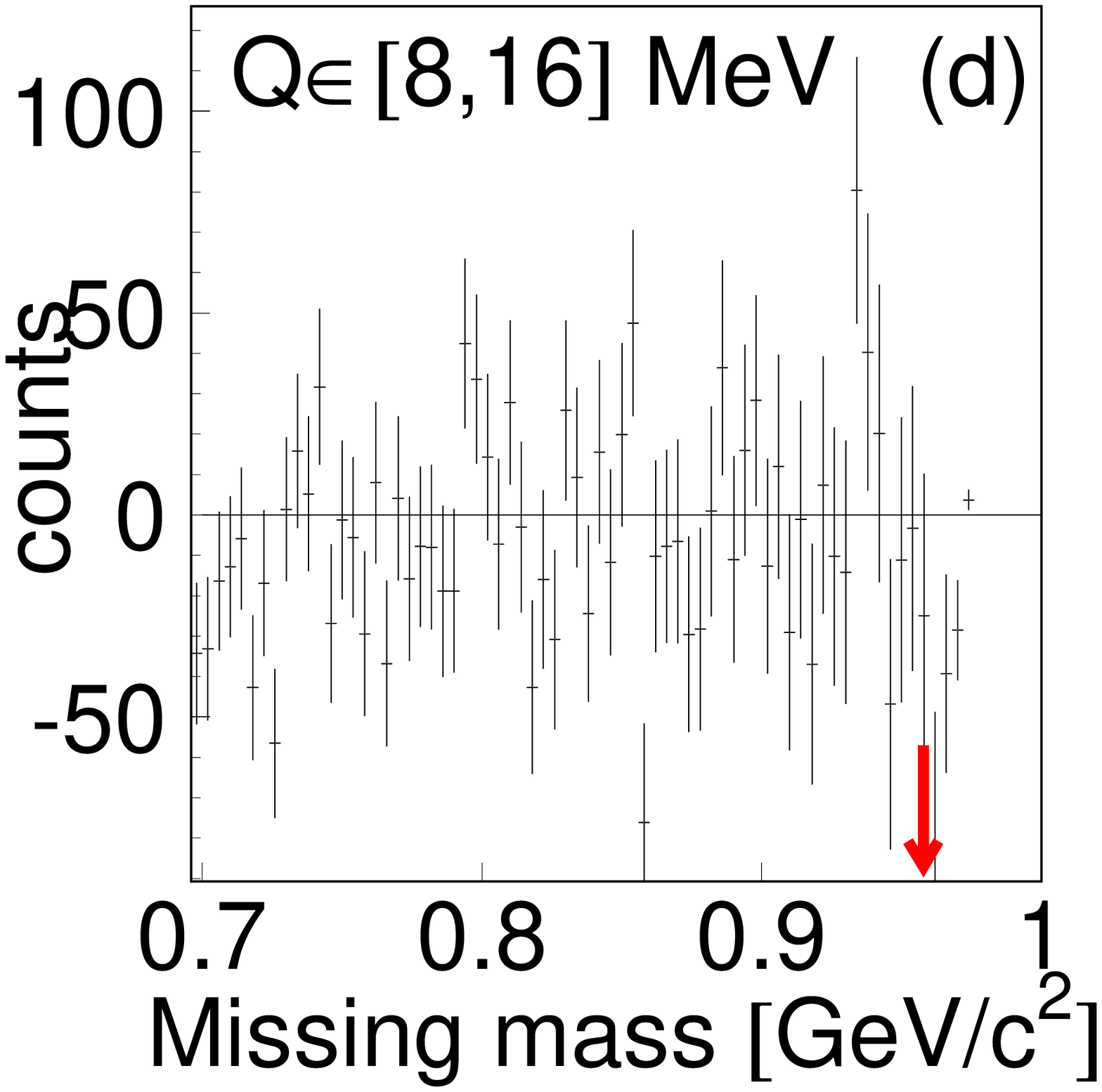}
\includegraphics[width=0.27\textwidth,angle=0]{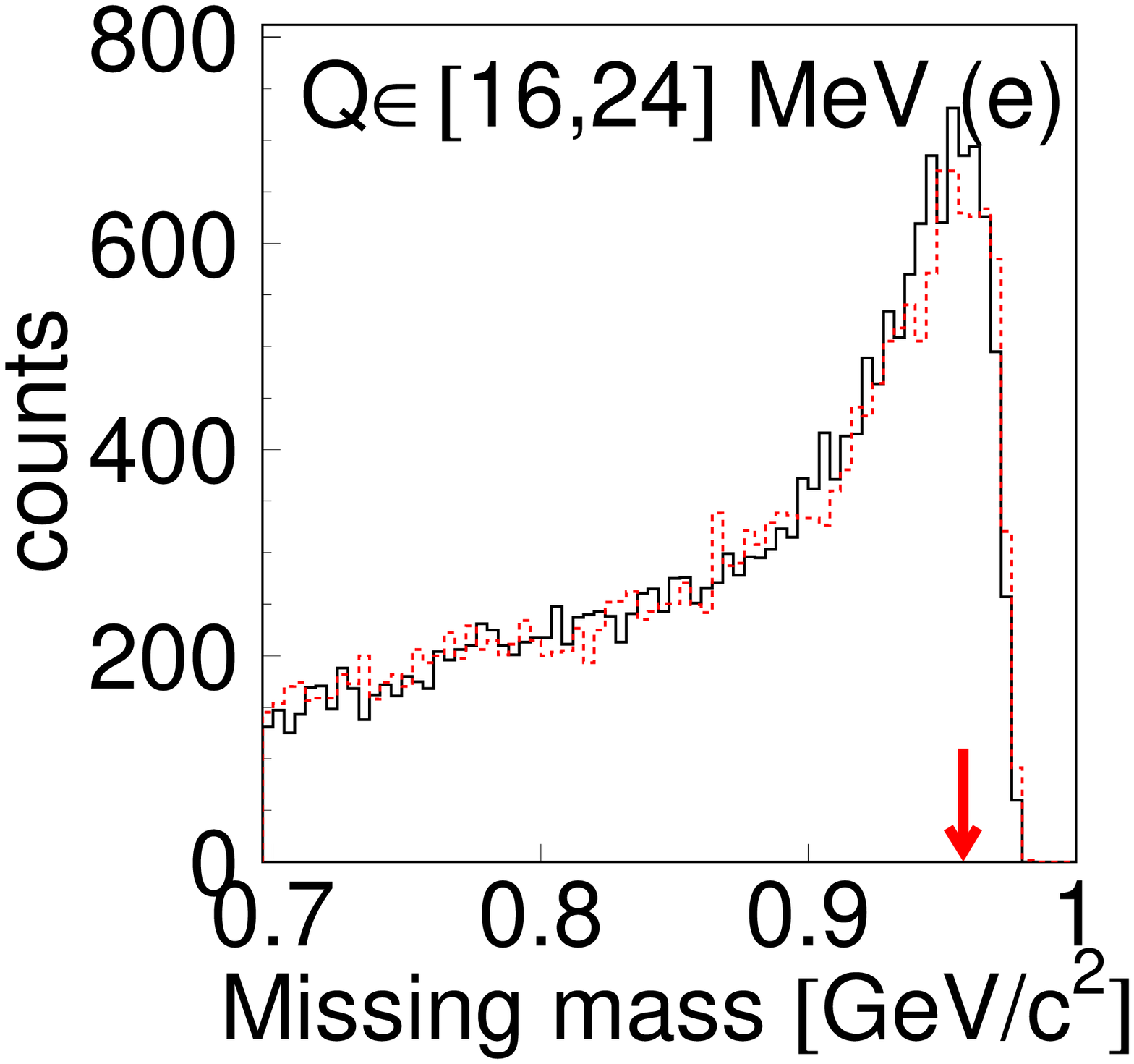}
\hspace{-1.25cm}
\includegraphics[width=0.27\textwidth,angle=0]{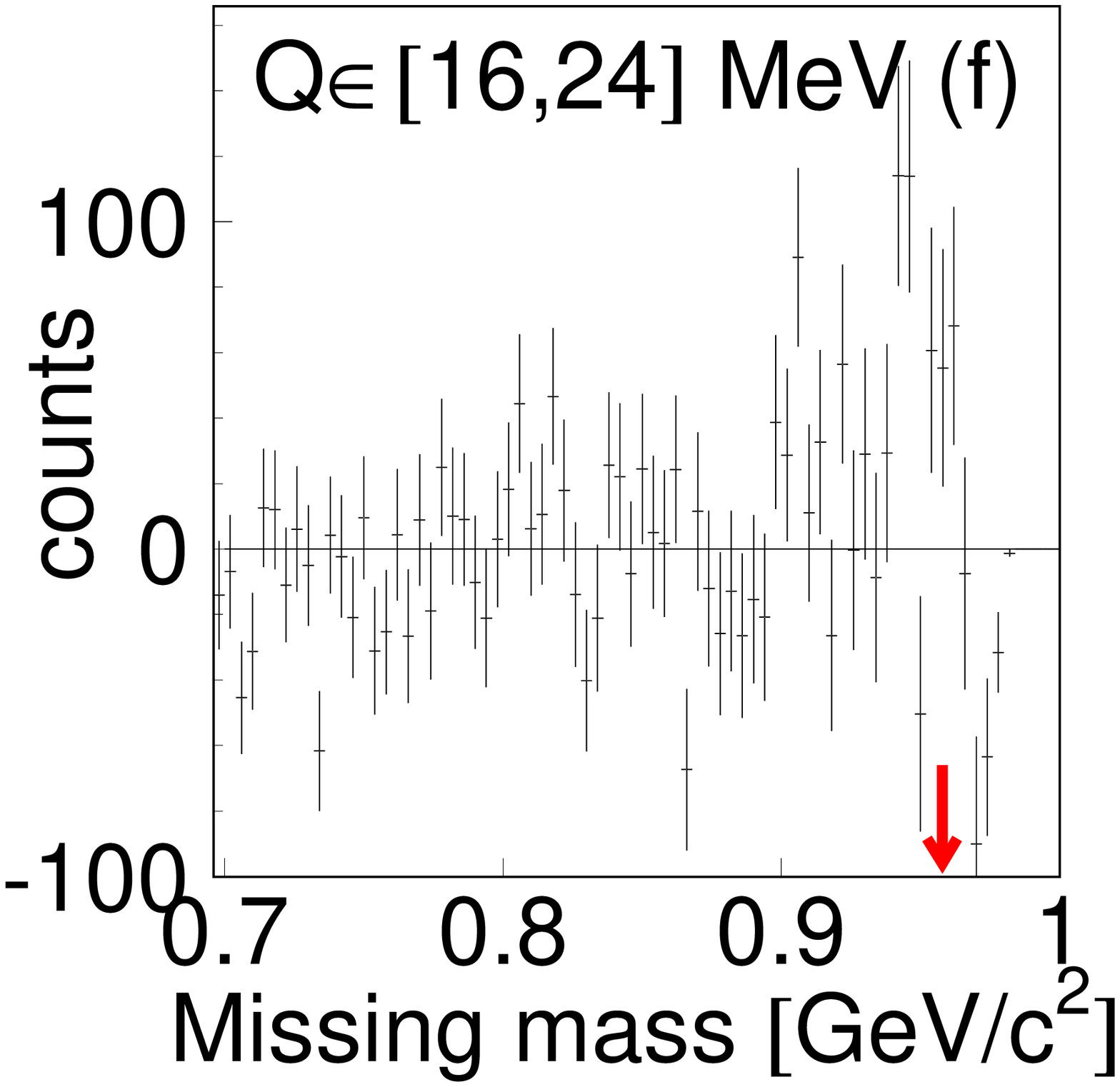}
\caption{{\bf (a, c, e)} 
Experimental missing mass distributions of the $pn \to pnX$ reaction 
(solid lines). Distributions were obtained for ranges of  
excess energy $Q$ from 0 to 8~MeV (a),
from 8 to 16~MeV (c) and from 16 to 24~MeV (e).
Corresponding background spectra are shown as dashed lines.
{\bf (b, d, f)} Experimental missing mass distributions of 
the $pn \to pn\eta^{\prime}$
reaction for excess energy ranges from 0 to 8~MeV (b),
from 8 to 16~MeV (d) and from 16 to 24~MeV (f)
determined after  background subtraction. 
Vertical bars indicate statistical
errors. The arrow depicts the nominal value of the $\eta^{\prime}$ mass.}
\label{mm_da_q0_8}
\end{figure}

For further analysis the collected data were grouped
according to the excess energy. 
The available range of excess energy $Q$
above the $\eta^{\prime}$-meson production threshold has been
divided into three 8~MeV-wide intervals. The choice of the width of the interval is a
compromise between statistics and resolution and reflects the accuracy
(FWHM) of the determination of the excess energy. 
In Figs~\ref{mm_da_q0_8}(a),~\ref{mm_da_q0_8}(c),~\ref{mm_da_q0_8}(e) 
the experimental mass distributions 
determined for the excess energy ranges 
$[0,8]$, $[8,16]$ and $[16,24]$~MeV are shown by solid lines.
For all distributions we observe an increase in counting rate toward
the kinematical limit. The shape is mostly caused by the detector acceptance.
A signal from the $\eta^{\prime}$-meson production is expected on top
of the continuous multipion mass distribution at the position 
corresponding to the $\eta^{\prime}$ mass (0.95778~GeV/$c^{2})$~\cite{amsler-pdg},
which is denoted by the arrow in Fig. 3.
However, no enhancement around this region is seen.

To extract a signal the histograms corresponding to  positive 
$Q$ intervals were compared to the background established from 
the data below the threshold.
For each distribution of the missing mass spectra at $Q \ge 0$, a corresponding
background spectrum was constructed from events with $Q$ 
belonging to the range $[-30,-10]$~MeV
by applying the method described in detail in a dedicated 
article~\cite{moskal-jpg32}.
This method was already applied successfully to the analysis of the 
$pn \to pn\eta$ 
reaction~\cite{moskal-prc79}.
As a cross-check, to gain more confidence in the applied procedure
we constructed alternative background distributions taking events 
from the $Q$ range from -20 to 0~MeV. 
The result of this investigation has shown a consistency of both 
background shapes of the order of 1\%.
Next, for each distribution for $Q \ge 0$ the corresponding 
background spectrum was normalised for mass values smaller than 
0.25~GeV/$c^2$
where no events from the $\eta^{\prime}$ meson are expected. 
In this region events correspond to one-pion production, for which 
the total cross section remains nearly constant in the range of 
the excess energy shown, because for a single-pion production 
the excess energy is high above the threshold~\cite{moskal-jpg32}.
The dashed lines in Fig.~\ref{mm_da_q0_8} depict the background 
spectrum, shifted to the kinematical limit and normalized accordingly.
The normalization in the  missing mass range significantly below
the mass of the 
$\eta^{\prime}$ meson (in the range from 0.75 to 0.85~GeV/$c^2$)
leads to the same result.

Figures~\ref{mm_da_q0_8}~(b),~\ref{mm_da_q0_8}(d),~\ref{mm_da_q0_8}(f) 
show distributions of the missing mass 
for the $pn \to pn\eta^{\prime}$ reaction as determined after 
subtraction of the background. 
Because of the low statistics and very low signal-to-background ratio 
the signal from $\eta^{\prime}$-meson created in the proton-neutron 
collision is statistically insignificant. 
Therefore, we can only estimate the upper limit for the
$\eta^{\prime}$ meson production in the $pn \to pn\eta^{\prime}$ reaction.

To estimate the systematic error of the number of 
 $pn \to pnX$ events,
the change in the missing mass spectra was studied by 
varying different parameters describing
the experimental conditions, discussed in the following
analysis and simulation~\cite{jklaja-phd}.
The change in the global time offset of the neutral particle detector by 1 
standard deviation of its time resolution 
$(\sigma_{t}^{N}~=~0.4$~ns) resulted in a systematic
error of the cross sections of 3\%. 
A change in simulations of the time resolution
of the neutron detector (0.4~ns)
by $\pm$0.2~ns resulted in $\pm$5\% changes in the cross section.
The variation in the cut on the noises in the spectator detector
within the limits of  energy resolution results in a change of cross sections by $\pm$7\%.
In the simulation we also checked  that varying the
beam momentum resolution arbitrarily by
$\pm$1~MeV/c around its nominal value of 3.35~MeV/c
resulted in changes in the cross section values by $\pm$3\%.
As discussed previously, and demonstrated in Ref.~\cite{moskal-jpg32}, 
the systematic error owing to the method used 
for the background subtraction is equal to 1\%.
The uncertainty owing to the model dependence calculation of the Fermi 
momentum distribution is equal to 2\%~\cite{czyzyk-dt} and
was estimated as the difference in results determined using the 
Paris~\cite{lacombe-plb101}
versus the CD-Bonn~\cite{machleidt-prc63} potentials. 
It is worth stressing that also a difference between the 
experimental Fermi momentum distribution and the predictions 
based on the Paris potential leads to a variation in the final result of about 2\%. 
A displacement of the spectator detector 
by 1~mm 
(the estimated accuracy of its position determination) 
changes the number of events by 5\%.
The error of the detection efficiency,
 predominantly caused by the 
uncertainty in the efficiency of the neutral particle detector~\cite{zdebik}, 
is determined to be no larger than 5\%~\cite{moskal-prc79}.
We also took into account  the systematic error 
($\sigma L_{int}~\approx~7\%$~\cite{jklaja-phd})
originating from the luminosity determination.
The  nuclear 
``shadow effect'' (estimated to about 4.5\%~\cite{chiavassa})
was not taken into account because
the reduction of the beam flux  seen by the neutron,
owing to the shielding by a spectator proton,
is expected to be the same for the $pn\to pn\eta^{\prime}$ reaction
and quasielastic scattering
that was used for  determination of the luminosity.
Similarly we have neglected the effect owing to the reabsorption
of the produced meson by the spectator proton, because even for the case of
$\eta$-meson production  
this effect reduces the cross section
by a factor of about 3\%~\cite{chiavassa2} and the proton-$\eta$ 
interaction is much stronger than the proton-$\eta^{\prime}$ interaction~\cite{swave}.
Finally, the total systematic error of $\sigma^{TOT}(Q)$ was estimated 
as the quadratic sum of the  nine independent systematic errors
discussed and is equal to $\approx~14\%$.

\section{Results}

\begin{figure}[h]
\centering 
\includegraphics[width=0.50\textwidth,angle=0]{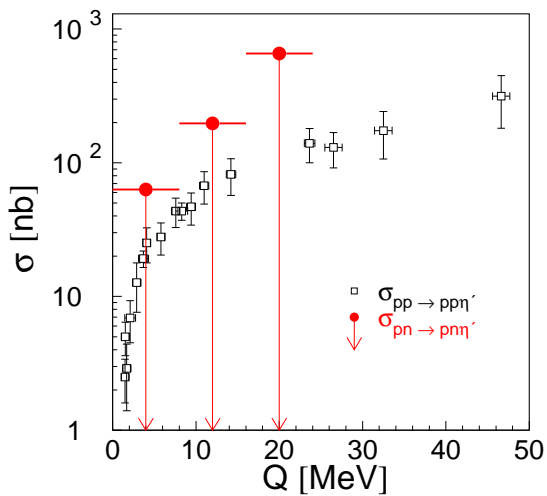}
\caption{ Total cross sections for the $pp \to pp\eta^{\prime}$
         reaction as a function of the excess energy (open squares).
         Upper limit for the total cross section for the $pn \to pn\eta^{\prime}$
         reaction as a function of the excess energy (filled circles).}
\label{cross_pn}
\end{figure}

The luminosity of $L~=~4.77~pb^{-1} \pm 0.06~pb^{-1}$
was established from the number of quasifree 
proton-proton scattering events applying the method described in 
the dedicated article~\cite{moskal-czyzyk-aip950}.
The acceptance of the detector setup and efficiency was determined 
based on Monte Carlo studies.
The signal from the $\eta^{\prime}$ meson in the determined missing mass 
spectra is statistically insignificant.
Therefore only the upper limit of the total cross section for the quasifree
$pn \to pn \eta^{\prime}$ reaction was extracted. 
The number of  $\eta^{\prime}$ mesons can be calculated as 
the difference between the number of events $(N^{SIG})$ in the 
peak for the signal 
(solid line in Fig.~\ref{mm_da_q0_8}) 
and the number of events $(N^{BACK})$ in the background of the 
peak (dashed line in Fig.~\ref{mm_da_q0_8}):
\begin{equation}
N^{\eta^{\prime}} = (N^{SIG} - N^{BACK}) \pm  \sqrt{N^{SIG}+N^{BACK}}.
\end{equation}
The range for the integration was chosen based on the simulation of 
the missing mass distributions~\cite{jklaja-phd}.
Assuming that  no $\eta^{\prime}$ mesons are observed 
$(N^{\eta^{\prime}}~=~0)$ in the experiment, 
the value of 1.2815 $\times$ $\sqrt{N^{SIG}+N^{BACK}}$ gives the upper 
limit of the total cross section for the $pn \to pn\eta^{\prime}$ 
reaction at a 90\% confidence level. 
The result is shown
in Fig.~\ref{cross_pn} and in Table~\ref{table_pn}.
\begin{table}[h]
\begin{center}
\begin{tabular}{|c|c|c|c|}
\hline
          & Upper limit of    &      & Upper limit of \\
$Q$ (MeV) & $\sigma(pn \to pn\eta^{\prime})$ & $\sigma(pp \to pp\eta^{\prime})$ (nb) & $R_{\eta^{\prime}}$ \\
          & at 90\% CL (nb)   &      & at 90\% CL\\
\hline
$[0,8]$   & 63  & 19.6 & 3.2\\
$[8,16]$  & 197 & 68.7 & 2.9\\
$[16,24]$ & 656 & 122.7 & 5.3\\
\hline
\end{tabular}
\end{center}
\caption{Excess energy range, upper limit of
         the total cross section for the $pn \to pn\eta^{\prime}$ reaction,
         mean value of the $pp \to pp\eta^{\prime}$ total cross section according to
         the parametrization given in Ref.~\cite{moskal-ijmpa22},
         upper limit of the ratio $R_{\eta^{\prime}}$ as a function of the excess energy.
         Energy intervals correspond to the binning applied.
         }
\label{table_pn}
\end{table}

The horizontal  bars in Fig.~\ref{cross_pn} represent the intervals of the excess 
energy for which the upper limit of the total cross section was calculated.

\begin{figure}[h]
\centering
\includegraphics[width=0.50\textwidth,angle=0]{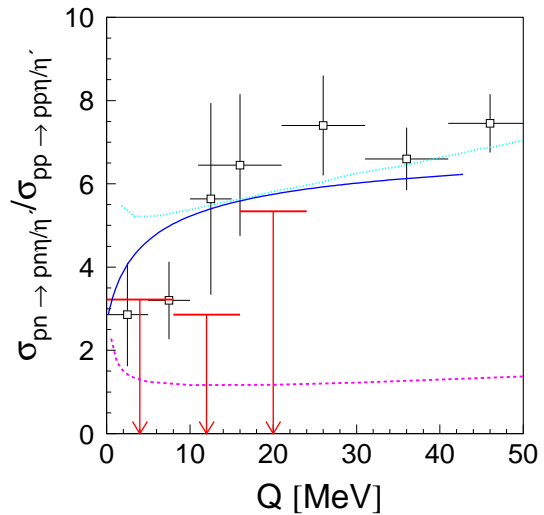}
\caption{Upper limit of the ratio $(R_{\eta^{\prime}})$ of 
         the total cross sections for the
         $pn \to pn\eta^{\prime}$ and $pp \to pp\eta^{\prime}$ reactions (arrows)
         in comparison with the ratio $(R_{\eta})$ determined for the $\eta$ 
         meson~\cite{calen-prc58,moskal-prc79} (open squares).
         The superimposed solid line indicates the result of a fit to the $R_{\eta}$ data
         taking into account the final-state interaction of 
         nucleons~\cite{moskal-prc79}.
         The dotted line presents the result of calculations performed 
         under assumption of dominance of the 
         S$_{11}$(1535) resonance in the production process~\cite{cao-prc78},
         and the dashed line denotes the result obtained within  a covariant effective meson-nucleon
         theory including meson and nucleon currents with
         nucleon resonances S$_{11}$(1650), P$_{11}$(1710) and P$_{13}$(1720)
         \cite{kampfer-ep}
         } 
\label{ratio_etap}
\end{figure}

The total cross section for the $pp \to pp\eta^{\prime}$ reaction was measured in 
previous experiments~\cite{daneppetaprim}. It reveals a strong excess energy dependence, 
especially very close to threshold. This dependence must be taken into account when comparing 
 the results for the $pn \to pn\eta^{\prime}$ reaction that were established for 8-MeV 
excess energy intervals. Therefore, for a given interval of excess energy, we have determined 
the mean value of the total cross section for $pp \to pp\eta^{\prime}$ reaction using the 
parametrization of F{\"a}ldt and Wilkin~\cite{wilkin-plb382,wilkin-prc56} fitted to the 
experimental data~\cite{moskal-ijmpa22}. Figure~\ref{ratio_etap} presents the upper limit 
of the ratio $R_{\eta^{\prime}}~=~{{\sigma(pn \to pn\eta^{\prime})} / {\sigma(pp \to pp\eta^{\prime})}}$ 
of the total cross section for the $pn \to pn\eta^{\prime}$ and $pp \to pp\eta^{\prime}$ reactions 
as a function of the excess energy (arrows). The corresponding values are listed in Table~\ref{table_pn}.
The corresponding ratios for the $\eta$ meson are also shown as open squares in Fig~\ref{ratio_etap}. 
For the $\eta$ meson, 
the value of $R_{\eta}$ is $\approx~$6.5 at excess energies larger 
than $\sim 16$~MeV~\cite{calen-prc58}, which  
suggests the dominance of isovector-meson exchange in the production mechanism. The  decrease in $R_{\eta}$ 
close to the threshold~\cite{moskal-prc79} may be explained by the different energy dependence of the 
proton-proton and proton-neutron final-state interactions~\cite{wilkin-priv}. 

For the $\eta^{\prime}$ meson the upper limit of the ratio for the excess energy range $[0,8]$~MeV 
is nearly equal to values of the ratio obtained for the $\eta$ meson, whereas for larger excess 
energy ranges, $[8,16]$ and $[16,24]$~MeV, the upper limits of the ratio are lower by about 
1 standard deviation each. The dotted curve in Fig.~\ref{ratio_etap} corresponds to the prediction 
for $R_{\eta^{\prime}}$ by Cao and Lee~\cite{cao-prc78} and exceeds the observed limit, showing the 
importance of other exchange currents in the $\eta^{\prime}$ production. The prediction of Kaptari and 
K{\"a}mpfer~\cite{kampfer-ep} is well within the upper bound. A smaller $R_{\eta^{\prime}}$ than $R_\eta$ 
is also consistent with a possible greater role for singlet currents in $\eta^{\prime}$ production than 
$\eta$ production. If there are important new dynamics in the $\eta^{\prime}$ production process relative 
to $\eta$ production, a key issue is the relative phase~\cite{Fald_Wilkin_Phys_Scripta} of possible additional 
exchanges compared to the isovector currents that dominate $\eta$ production. The observed limit thus 
constrains modeling of the production processes. To confirm these interesting observations and to go farther, 
new experimental investigations with improved statistics are required.

\section{Summary}

To determine the excitation function of the total cross section for the quasifree $pn \to pn\eta^{\prime}$ 
reaction near the kinematical threshold, we performed an experiment at the cooler synchrotron
COSY at the Research Centre J{\"u}lich using the COSY-11 detector system. The quasifree $pn \to pn\eta^{\prime}$ 
reaction was induced by a proton beam with a momentum of 3.35~GeV/c on a deuteron target. All outgoing nucleons 
have been registered, whereas for $\eta^{\prime}$-meson identification the missing mass technique was applied.
The upper limit of the total cross section for the $pn \to pn\eta^{\prime}$ reaction in the excess energy range 
between 0 and 24~MeV has been determined. The total cross section for the $pp \to pp\eta^{\prime}$ reaction was 
measured in previous experiments at the same energy range. Combining these data, we have determined the upper limit of 
the ratio $R_{\eta^{\prime}}=\sigma(pn \to pn\eta^{\prime})/\sigma(pp \to pp\eta^{\prime})$. The comparison of the 
$R_{\eta}$ ratio with the upper limits of $R_{\eta^{\prime}}$ established here suggests  nonidentical production 
mechanisms for the $\eta^{\prime}$ and $\eta$ mesons, which are dominated by a strong isovector exchange contribution. 
To pin down the detailed reaction mechanism further theoretical as well as  experimental investigations with better 
statistics are required.

\begin{acknowledgments}

We are grateful for useful comments of Burkhard~K{\"a}mpfer,
Piotr Salabura and Joanna Stepaniak.
The work was partially supported by the European Community-Research 
Infrastructure Activity under the FP6 and FP7 programs 
(Hadron Physics, RII3-CT-2004-506078, PrimeNet No. 227431), 
by the Polish Ministry of Science and Higher Education under Grant
Nos 3240/H03/2006/31, 1202/DFG/2007/03, and 0082/B/H03/2008/34,
by the German Research Foundation (DFG),
by the FFE grants from the Research Center J{\"u}lich, 
by the Austrian Science Fund, FWF, through grant P20436
and by the virtual institute "Spin and strong QCD" (VH-VP-231).
 
\end{acknowledgments}

\end{document}